\documentclass[aps,prb,twocolumn,showpacs,floatfix]{revtex4}

\usepackage{graphicx}
\usepackage[squaren,thinspace,textstyle]{SIunits}
\addunit{\evolt}{e\volt}

\usepackage{color}

\newcommand{\om}{\Omega^2_\text{exc}}

\begin{document}

\title{Single exciton emission from gate-defined quantum dots}
\textcolor{blue}{}

\author{}
\affiliation{}

\author{G. J. Schinner}
\affiliation{Center for NanoScience and Fakult\"at f\"ur Physik,
Ludwig-Maximilians-Universit\"at,
Geschwister-Scholl-Platz 1, 80539 M\"unchen, Germany}

\author{J. Repp}
\affiliation{Center for NanoScience and Fakult\"at f\"ur Physik,
Ludwig-Maximilians-Universit\"at,
Geschwister-Scholl-Platz 1, 80539 M\"unchen, Germany}

\author{E. Schubert}
\affiliation{Center for NanoScience and Fakult\"at f\"ur Physik,
Ludwig-Maximilians-Universit\"at,
Geschwister-Scholl-Platz 1, 80539 M\"unchen, Germany}

\author{A. K. Rai}
\affiliation{Angewandte Festk\"orperphysik, Ruhr-Universit\"at Bochum, Universit\"atsstra{\ss}e 150, 44780 Bochum, Germany}

\author{D. Reuter}
\affiliation{Angewandte Festk\"orperphysik, Ruhr-Universit\"at Bochum, Universit\"atsstra{\ss}e 150, 44780 Bochum, Germany}

\author{A. D. Wieck}
\affiliation{Angewandte Festk\"orperphysik, Ruhr-Universit\"at Bochum, Universit\"atsstra{\ss}e 150, 44780 Bochum, Germany}

\author{A. O. Govorov}
\affiliation{Center for NanoScience and Fakult\"at f\"ur Physik,
Ludwig-Maximilians-Universit\"at,
Geschwister-Scholl-Platz 1, 80539 M\"unchen, Germany}
\affiliation{Department of Physics and Astronomy, Ohio University, Athens, Ohio 45701}

\author{ A. W. Holleitner}
\affiliation{Walter Schottky Institut and Physik-Department, Am Coulombwall 4a, Technische Universit\"at M\"unchen, D-85748 Garching, Germany}

\author{J. P. Kotthaus}
\affiliation{Center for NanoScience and Fakult\"at f\"ur Physik,
Ludwig-Maximilians-Universit\"at,
Geschwister-Scholl-Platz 1, 80539 M\"unchen, Germany}

\date{\today}

\begin{abstract}
With gate-defined electrostatic traps fabricated on a double quantum well we are able to realize an optically active and voltage-tunable quantum dot confining individual, long-living, spatially indirect excitons. We study the transition from multi excitons down to a single indirect exciton. In the few exciton regime, we observe discrete emission lines reflecting the interplay of dipolar interexcitonic repulsion and spatial quantization. The quantum dot states are tunable by gate voltage and employing a magnetic field results in a diamagnetic shift. The scheme introduces a new gate-defined platform for creating and controlling optically active quantum dots and opens the route to lithographically defined coupled quantum dot arrays with tunable in-plane coupling and voltage-controlled optical properties of single charge and spin states.
\end{abstract}


\maketitle

\section{Introduction}

Trapping atoms or ions with tailored electromagnetic fields as on microfabricated chips \cite{2010Riedel, 2008Blatt} or in optical lattices \cite{2011Weitenberg} and thus controlling their position and number down to single particles has enabled detailed studies of quantum systems such as Bose Einstein condensates \cite{2010Riedel}, coupled ion arrays \cite{2008Blatt} and atomic Mott insulators \cite{2011Weitenberg}. This allowed particle entanglement  \cite{2010Riedel, 2008Blatt} and yielded insight into light matter-interactions with individual quantum objects with unprecedented precision. In solid state devices, similarly, single electron manipulation in charge quantum dots \cite{2007Hanson, 2002Holleitner, 2007Marcus, 2011Hermelin} or Bose-Einstein condensation of two-dimensional bosonic systems in equilibrium both require a trapping potential \cite{1967Hohenberg}. So far, optically active quantum dots, trapping few excitons, rely on three dimensional material modulation as in self-assembled quantum dots \cite{1994Drexler, 2000Warburton_Natur, 2011Latta}, providing limited control of confinement potential and position. Here, we demonstrate in tuneable nanoscale traps that tight electrostatic confinement of excitons creates discrete excitonic transitions observable down to a single confined exciton.
Such traps combine a double quantum well with gate-defined electrostatic potentials resulting in individual three-dimensional traps for spatially indirect dipolar excitons \cite{2011Schinner}. With their electron and hole confined to a different quantum well, these excitons exhibit a large dipole moment and long lifetimes.
Reducing the electrostatic confinement to nanoscale dimensions we enter the few exciton regime and observe discrete spectral features, reflecting interexcitonic dipolar repulsion to cause molecular-like spatial arrangements of two and three excitons, respectively.
The excitonic transitions are tuneable by gate voltages and magnetic fields and are well reproduced by a straightforward model. The scheme introduces a new gate-defined platform for creating and controlling single exciton traps and opens the route to lithographically defined arrays of artificial atoms with tuneable in-plane coupling and voltage-controlled optical properties of single charge and spin states.

In solids quantum confinement of charge carriers in all three spatial directions results in so-called quantum dots (QD), artificial atoms with discrete energy spectra. Charge quantum dots, containing unipolar charges and studied via electronic transport spectroscopy, usually utilize electrostatic fields to completely confine charge carriers of a two-dimensional (2D) electron system at a heterojunction interface \cite{2007Hanson, 2002Holleitner, 2011Hermelin} or in an one-dimensional (1D) nanowire \cite{2007Marcus, 2011Hermelin}. In contrast, optically active quantum dots, containing both conduction band electrons and valence band holes so far require three-dimensional material modulation by suitable growth methods as in widely studied self-assembled quantum dots \cite{1994Drexler, 2000Warburton_Natur, 2011Latta} in order to confine both, electrons and holes, in the same nanoscale region as bound excitonic pairs. Thereby it is difficult to control their position to achieve scalable QD circuits and their confining potential. However, introducing charge tuneability via field-effect has enabled to visualize shell structure of single electron states and precision optical spectroscopy on individual self-assembled quantum dots in emission \cite{2000Warburton_Natur} and absorption \cite{2004Hogele} and continues to reveal new many-body quantum phenomena \cite{2011Latta}.

In an effort to combine the advantages of the above approaches we present a scheme to define optically active quantum traps confining individual indirect dipolar excitons (IX) by electrostatic fields provided via lithographically fabricated nano-scale gates on a suitably designed double quantum well (DQW) heterostructure. By reducing the trap area and the corresponding IX occupation about 1000-fold in comparison to previous studies \cite{2011Schinner, 2009HighNanoL} we achieve the single exciton limit for electrostatically trapped and long-living dipolar excitons. We thus are able to observe discrete emission lines for the 1\,IX, 2\,IX and 3\,IX states reflecting quantum dot behavior dominated by the effect of dipolar repulsion between individual excitons, in excellent agreement with a model of interacting IX in a tight parabolic confinement potential. This enables position control of such QDs with lithographic precision as in charge and spin QDs. In addition, it permits electrical tuneability of the excitonic confining potential, photoluminescence (PL) energy and lifetime, and the exciton population down to a single exciton. Based on our approach one can envision the control of individual excitons in the DQW plane within QDs or in QD arrays with gate-controlled in-plane coupling as needed for implementation in quantum information processing circuits  \cite{1998Loss, 1999Imamog, 1999Burkard, 2003Li}. Furthermore IX are particularly attractive for coherent manipulation, because of their long charge and spin lifetimes \cite{2009Vörös, 2010Kowalik-Seidl-spin}.

\section{The quantum traps}

\begin{figure}[h]
\centering
\includegraphics[width=8.6cm, angle=0]{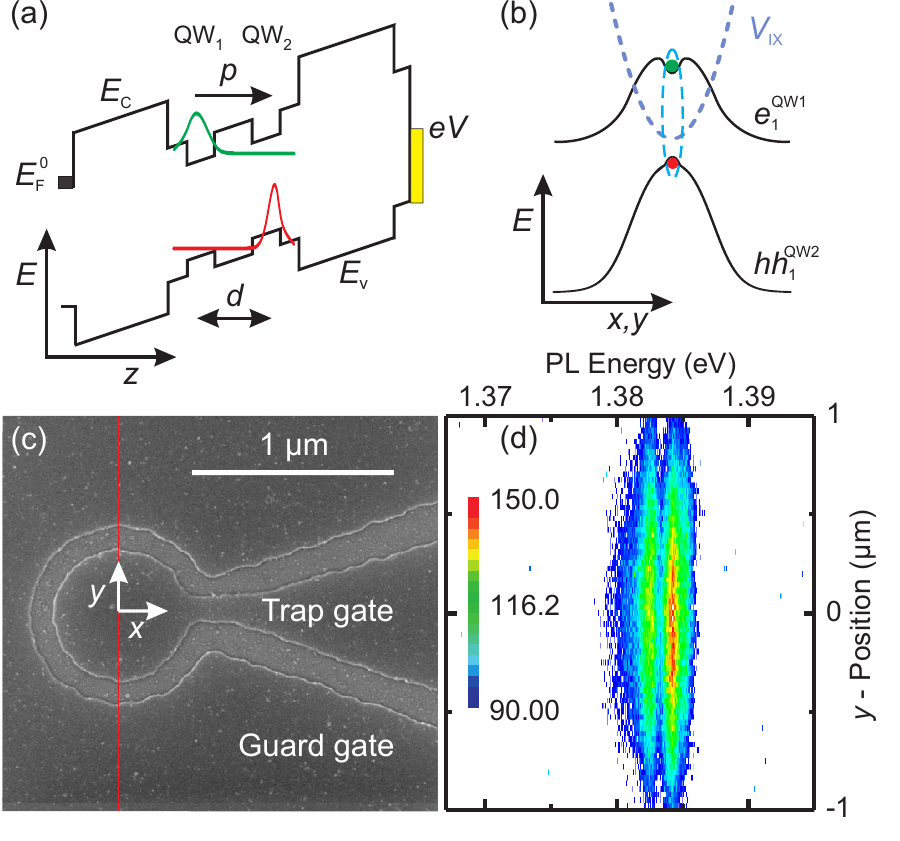}
\caption{\label{QDfig1} \textbf{Trapping principle.} (a) depicts a schematic of the DQW band gap within a field-effect device. $E_{\mathrm{c}}$ and $E_{\mathrm{v}}$ denote the conduction and valence band edge, respectively. The electron and hole wavefunction of the ground states $hh_{1}^{\mathrm{QW2}}$ and $e_{1}^{\mathrm{QW1}}$ are indicated. Note that flat band occurs at $V_{\mathrm{T}} \simeq V_{\mathrm{G}}\simeq$ 0.8\,V \cite{2011Schinner}. (b) in-plane energy of the $hh_{1}^{\mathrm{QW2}}$ and $e_{1}^{\mathrm{QW1}}$ ground state and the resulting confinement potential $V_{\mathrm{IX}}$ for the IX in the quantum dot. (c) displays a scanning electron microscope (SEM) image of the gate pattern defining the quantum dot via the trap and guard gate. In (d) is shown the PL intensity in a logarithmic color scale as a function of the PL energy and $y$-position along a cut through the middle of the trap as indicated by the red line in (c) [$V_{\mathrm{G}}$ = 0.75\,V, $V_{\mathrm{T}}$ = 0.25\,V, $T_{\mathrm{Lattice}}$ = 242\,mK, $P_{\mathrm{Laser}}$ = 2.8\,nW, $E_{\mathrm{Laser}}$ = 1.494\,eV].}
\end{figure}

The 3D confining scheme in the IX quantum traps is based on gate control of the Quantum Confined Stark Effect (QCSE) in the plane of a DQW as illustrated in Fig.\ \ref{QDfig1}. The electric field component $\mathbf{F}_{\mathrm{z}}$ in the growth direction ($z$) of the coupled DQW consisting of two adjacent 7\,nm wide $\mathrm{In}_{0.11}\mathrm{Ga}_{0.89}\mathrm{As}$  QWs separated by center to center distance of $d$ = 17\,nm is tuned by a voltage $V$ applied to semi-transparent titanium gates with respect to a n-doped GaAs back contact with Fermi-energy $E_{\mathrm{F}}^{0}$ as shown in the energy diagram in Fig.\ \ref{QDfig1}(a). The e-h separation causes a red shift $\Delta E_{\mathrm{IX}}=-\mathbf{pF}_{\mathrm{z}}$ of the IX energy where $\mathbf{p}=e\mathbf{d}$ is the IX dipole moment oriented along $z$.
The parabolic confinement $V_{\mathrm{IX}}$ in the $x-y$-plane (Fig.\ \ref{QDfig1}(b)) is achieved by biasing the trap gate (Fig.\ \ref{QDfig1}(c)) with a diameter of 600\,nm at a negative voltage $V_{\mathrm{T}}$ with respect to the guard gate. The latter is biased close to the flat-band voltage $V_{\mathrm{FB}}$ at which the built-in field caused by surface states is canceled by the gate voltage.  Coulomb attraction by the hole, localized beneath the center of the trap gate, binds the electron with a typical binding energy of 3\,meV and prevents excitonic ionization by the external in-plane electric field. The spatial separation of the electron and hole along $z$ reduces the overlap of the corresponding wavefunctions (Fig.\ \ref{QDfig1}(a)) resulting in a radiative lifetime of order 100\,ns for the IX (see Section \ref{LifetimeQD}). Filling the QD with an increasing number of individual IXs allows the investigation of the interplay between dipolar IX-IX repulsion and spatial quantization in the QD. In addition, raising the number of hydrogen-like bosonic IXs, one enters the interesting regime, where excitonic Bose-Einstein condensation (BEC) is expected as predicted in the1960s \cite{1962Blatt}. Recently, micrometer scale electrostatic trapping configurations were reported to realize a cooled ensemble
of indirect excitons with high densities in order to implement a BEC (see Section \ref{Theelectrostatictrap}) \cite{2011Schinner, 2011Schinner-correlation}.

The devices are investigated in a low temperature microscope with two diffraction limited confocal objectives that can be independently positioned in a $^{\mathrm{3}}$He refrigerator at 240\,mK. Using the upper objective, we create IXs under the guard gate with laser light 1.2\,$\mu$m away from the trap center. The quantum trap is thus only filled with IXs pre-cooled to lattice temperatures, whereas free electrons are unable to enter the trap \cite{2011Schinner}. The emitted PL light is collected in transmission with a second objective located below the device and analyzed in a spectrometer. An image of the PL intensity (Fig.\ \ref{QDfig1}(d)) shows two discrete energy-resolved lines, emitted from a single exciton (1\,IX) and a biexciton (2\,IX) as discussed below, located in the center of the trap. They are spatially diffraction broadened to 900\,nm full width at half maximum, comparable to the emitted PL wavelength.

\section{The quantum dot trap populated with few excitons}

\begin{figure}[h]
\centering
\includegraphics[width=8.6cm, angle=0]{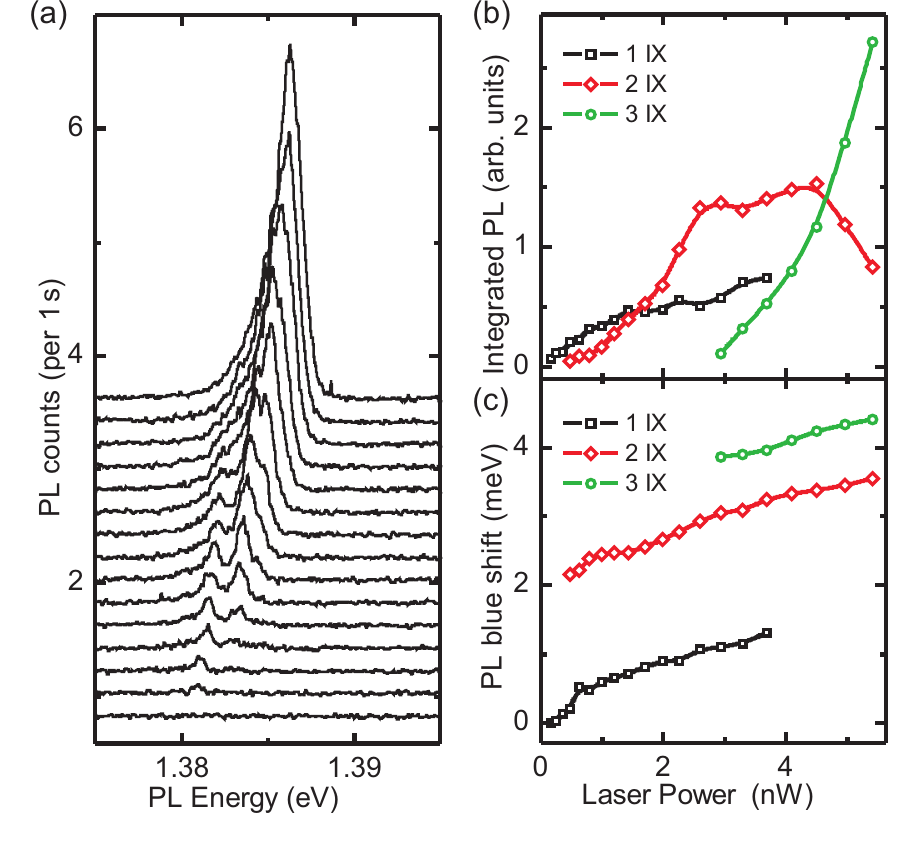}
\caption{\label{QDfig12} \textbf{Laser power dependence.} (a) illustrates typical PL spectra for different occupations of the quantum dot nonlinearly changed by laser power between 55\,pW and 10\,nW. In comparison to the emission from individual self-assembled InGaAs quantum dots PL counts in the single exciton limit are typically found to be 1000-fold lower, reflecting the correspondingly increased radiative lifetime of IX. The corresponding integrated PL intensity and PL blue shift is shown in (b) and (c) [$V_{\mathrm{G}}$ = 0.65\,V, $V_{\mathrm{T}}$ = 0.30\,V, $T_{\mathrm{Lattice}}$ = 245\,mK, $E_{\mathrm{Laser}}$ = 1.494\,eV].}
\end{figure}

The population of the QD can be varied by changing the laser power exciting IXs in the vicinity of the QD (Fig.\ \ref{QDfig12}(a)). For excitation laser powers up to 300\,pW incident on the sample surface we observe a single narrow PL line, which we associate with an individual IX. Increasing the laser power generates two discrete PL lines (see also Fig.\ \ref{QDfig1}(d)), demonstrating the occupation of the QD with a biexciton decaying via a single IX. A further increase in excitation power creates in addition triexcitons and so on. The discrete PL lines are narrow ($\sim$0.4 meV) but orders of magnitude broader than expected from the radiative lifetime of 300\,ns. This we attribute to temporal spectral fluctuations happening during the long integration times of order 200 seconds necessary for detection. The dependence of the PL intensity on the laser power shown in (Fig.\ \ref{QDfig12}(b)) is characteristic for a QD populated with multiexcitons \cite{1994Brunner}, rising e.g.\ quadratically in laser power for a biexciton.
The center energies of the discrete PL lines, plotted in (Fig.\ \ref{QDfig12}(c)), reflect the interplay between dipolar interexcitonic repulsion and spatial quantization leading to a splitting of nearly 2\,meV between the single exciton and biexciton configuration.

\begin{figure*}[htb]
\centering
\includegraphics[width=\textwidth, angle=0]{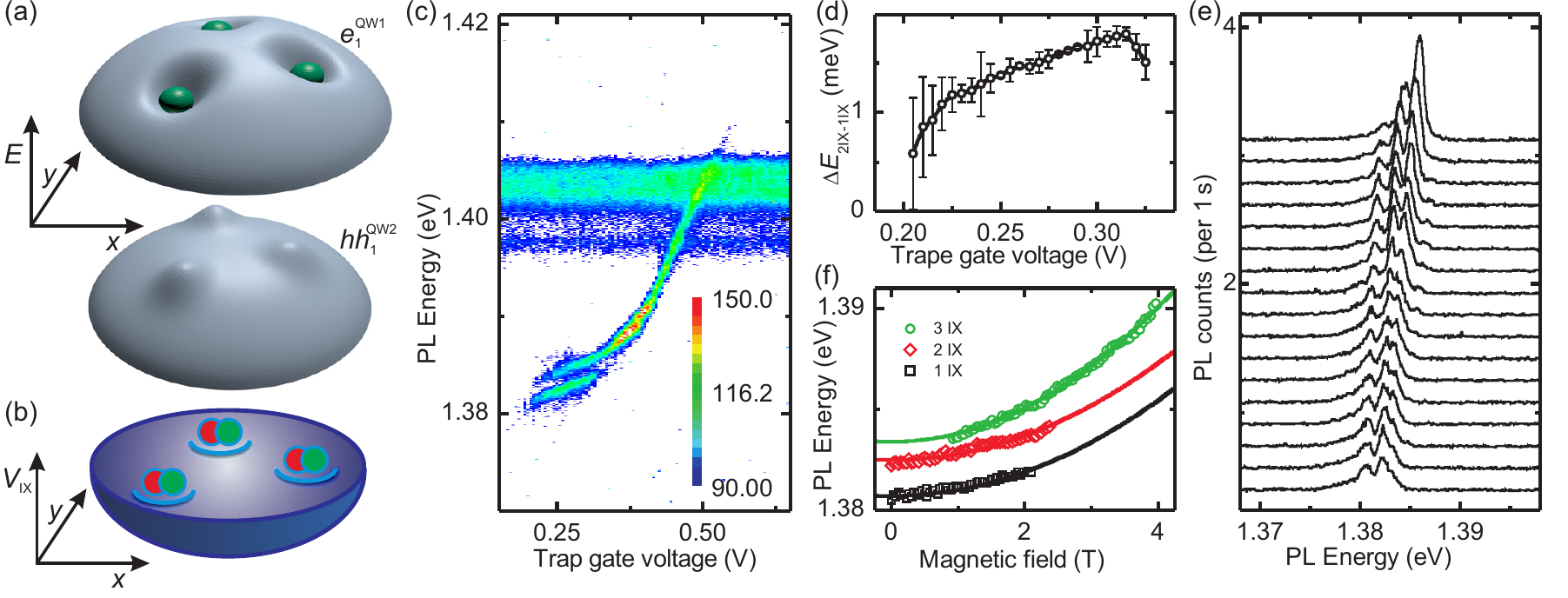}
\caption{\label{QDfig2} \textbf{Wigner-molecule, trap gate and magnetic field dependence.} (a) shows the formation of a triangular excitonic Wigner-molecule built up of three equidistantly spaced dipolar excitons confined in the QD like in Fig.\ \ref{QDfig1}(b). (b) illustrates the corresponding confinement potential $V_{\mathrm{IX}}$ containing 3\,IX. (c) depicts the PL intensity in a logarithmic color scale as a function of the trap gate voltage and the PL energy. With decreasing trap gate voltage the PL line splits in two. This splitting as a function of the trap gate voltage is shown in (d) [$V_{\mathrm{G}}$ = 0.55\,V, $T_{\mathrm{Lattice}}$ = 244\,mK, $P_{\mathrm{Laser}}$ = 2.8\,nW, $E_{\mathrm{Laser}}$ = 1.494\,eV]. (e) shows characteristic spectra for different magnetic fields in equidistant steps from 0\,T (bottom) to 2.5\,T (top). The corresponding diamagnetic shifts of the different PL lines (symbols) and the quadratic fit (solid line) are shown in (f). From the quadratic fit we can extract an effective Bohr radius $r_{\mathrm{e}}$ ($r_{\mathrm{e}}^{1\mathrm{IX}}=18.3$\,nm, $r_{\mathrm{e}}^{2\mathrm{IX}} = 18.4$\,nm and $r_{\mathrm{e}}^{3\mathrm{IX}} = 21.7$\,nm) [$V_{\mathrm{G}}$ = 0.75\,V, $V_{\mathrm{T}}$ = 0.25\,V, $T_{\mathrm{Lattice}}$ = 242\,mK, $P_{\mathrm{Laser}}$ = 2.8\,nW, $E_{\mathrm{Laser}}$ = 1.494\,eV].}
\end{figure*}

Based on a semiclassical model (see Section \ref{ModelQD}), we can describe the level structure by the following scenario.
With the heavy hole fixed in space by the gate-induced electrostatic confining potential, somewhat roughened by the random disorder potential, the light electron in the adjacent QW is electrostatically bound to the hole via Coulomb attraction (Fig.\ \ref{QDfig1}(b)). Filling an additional IX into the narrow lateral confining potential of the QD trap causes strong excitonic repulsion by dipolar interaction. The repulsive energy of about 2\,meV at a lateral interexcitonic distance of about 34\,nm (see Section \ref{ModelQD}) dominates the energy splitting between the single IX and biexciton. Adding the third exciton yields a somewhat reduced additional repulsion of about 0.7\,meV at a distance of 37\,nm if we assume a triangular ordering of the excitons in the trap (see Section \ref{ModelQD}).
The strong dipolar interactions between the few excitons in the quantum dot cause spatial order of the excitons \cite{2009Laikhtman}, in analogy to Wigner-crystal-like states expected in charge quantum dots \cite{1994Peeters}. This spatial order results in an excitonic Wigner-molecule (Fig.\ \ref{QDfig2}(a), (b)) and is likely to be assisted by the fact that the external excitonic potential $V_{\mathrm{IX}}$ has a significantly larger spatial extend than the excitons confined to Bohr radii.
In addition, screening of the bare trap potential $V_{\mathrm{IX}}$ by adding dipolar excitons will modify the effective potential landscape.
Increasing the laser power increases the time-averaged population of the trap with biexcitons and triexcitons. The resulting dipolar screening causes a blue shift $\Delta E$, reflecting the increasing many-body interactions between the IXs and their temporal dynamics (Fig.\ \ref{QDfig12}(c)). A further increase of the laser power causes the QD to sequentially fill up to about 100\,IXs (see Section \ref{populatedQD}). For high populations we observe an unstructured asymmetric PL lineshape characterized by a steep blue side and a long red tail (Fig.\ \ref{QDfig12}(a)). The development of the PL lineshape in Fig.\ \ref{QDfig12}(a) with the trapped exciton density shows clearly that the asymmetric PL line results from the sum of individual exciton PL lines and ends in an edge-like singularity \cite{1987Skolnick, 1991Hawrylak} for a high density cold dipolar liquid \cite{2011Schinner-correlation, 2009Laikhtman}.

In the quantum trap device we anticipate to observe correlations because the expected excitonic coherence length \cite{2006Yang} exceeds the diameter of the trap which is also smaller than the emitted PL wavelength. Furthermore, at low temperature (240\,mK), the thermal de Broglie wavelength $\lambda_{\mathrm{dB}}$ of the bosonic particles ($\lambda_{\mathrm{dB}} = \frac{\hbar}{\sqrt{2M\pi kT}}$ where $M=m_{\mathrm{e}}^{*}+m_{\mathrm{hh}}^{*}$ is the total mass of the exciton) is a few times larger than the interexcitonic distance which is similar to the effective Bohr radius $r_{\mathrm{e}} \approx$ 20\,nm.

The excitonic QD potential $V_{\mathrm{IX}}$ is tuneable by gate voltage (Fig.\ \ref{QDfig2}(c)). With decreasing trap gate voltage, the trap gets deeper and we expect a narrowing of the lateral confinement down to about 100\,nm, inducing a splitting of the PL line in two discrete lines (see Section \ref{ConfinementQD}). The latter reflects an occupation of the trap with two exciton states. In Fig.\ \ref{QDfig2}(d), the energetic splitting between exciton and biexciton state is plotted versus the trap gate voltage. We interpret the lowering of the energetic spacing $E_{\mathrm{2IX}}-E_{\mathrm{1IX}}$ with decreasing trap gate voltage by the observation that the time-averaged excitonic population of the quantum trap with biexcitons, deduced from the decreasing intensity of the 2\,IX PL in Fig.\ \ref{QDfig2}(c), and thus the corresponding effective excitonic repulsion weakens and overcompensates the effect of a somewhat increasing electrostatic potential $V_{\mathrm{IX}}$. The splitting is accompanied by a change in the slope of the QCSE (Fig.\ \ref{QDfig2}(c)). Here, we insured that leakage currents entering the trap gate are negligibly small in the corresponding voltage regime (see Section \ref{Leakage currents}).
In traps with a larger trap gate diameter we are able to observe a gate-voltage-dependent depletion zone of width ranging between 250\,nm and 500\,nm around the trap perimeter (see Section \ref{ConfinementQD}) \cite{2011Schinner}. Correspondingly, we expect for the trap depicted in Fig.\ \ref{QDfig1}(c) the resulting curvature of the electrostatically induced trap potential $V_{\mathrm{IX}}$ to dominate the random disorder potential. Generally, at a trap gate voltage $V_{\mathrm{T}} \gtrsim V_{\mathrm{G}}$, the exciton trap converts into a depopulated antitrap and we do no longer detect any PL from IXs within the trap area.

Additional manifestation of the quantum dot behavior can be deduced from the difference in the diamagnetic energy shifts of the discrete excitonic lines in a magnetic field $B$ applied in Faraday configuration perpendicular to the QW plane. Fig.\ \ref{QDfig2}(e) displays exemplary PL spectra as a function of $\mathbf{B}$. With increasing magnetic field we observe a diamagnetic shift of the excitonic emission with the center energies plotted in Fig.\ \ref{QDfig2}(f). In addition, we observe an increasing PL intensity and the appearance of a triexciton line at higher field. Both findings reflect an increasing stabilization of excitons by the magnetic field. Without magnetic field, a large in-plane component of the electrostatic field near the trap perimeter can cause ionization of the IX. As experimentally observed inside QWs \cite{1997Zimmermann} and theoretically predicted \cite{1998Govorov}, such radial tunneling of the excitonically bound electrons out of their pocket in the binding potential (Fig.\ \ref{QDfig1}(b) and Fig.\ \ref{QDfig2}(a)) can be suppressed by a magnetic field along $z$. For the spectra in Fig.\ \ref{QDfig2}(e), this causes both the observed increase in PL intensity and the appearance of a strong triexciton line with increasing $\mathbf{B}$.
The resulting quadratic diamagnetic shift of the trapped IXs energies (plotted in Fig.\ \ref{QDfig2}(f)) is given by $\Delta E _{\mathrm{dia}}= \frac{e^2 r_{\mathrm{e}}^2}{8 \mu c^2}B^2$ where $r_{\mathrm{e}}$ is the effective Bohr radius and $\mu$ the reduced mass of the exciton \cite{2008SternMott}.
With increasing exciton number, we see an increase in $ r_{\mathrm{e}}$ because the effective confining potential is broadened by screening and strong IX interactions.

\section{\label{Theelectrostatictrap}The electrostatic trapping principle of the quantum traps}

To create electrostatic traps we use the in-plane variation of the local Stark electric field $\mathbf{F}_{\mathrm{z}}(x,y)$ acting perpendicular to the double quantum well (DQW) in order to generate a potential landscape for the indirect excitons (IXs). Such highly tuneable traps were realized in various geometries \cite{1997Zimmermann, 1998zimmermann_APL, 1998Huber, 2005Rapaport, 2007Timofeev, 2006hammackTrap, 2006Chen, 2009HighNanoL, 2009HighDiamond, 2011Schinner}. Other possibilities to realize three-dimensional exciton confinement are: strain-induced traps \cite{1983Trauernicht, 1988Kash, 1999negoita}, magnetic traps \cite{1998Christianen}, natural traps due to random disorder in the QW-plane \cite{2002Butov_N} or silicon-dioxide traps \cite{2007Gartner, 2009Vogele}. Electrostatic traps exhibit the highest tunability and trapping efficiency. To achieve a quantitative understanding in the functionality of our gate-defined traps we investigated a series of circular traps with gate diameters of the central trap gate ranging from 24\,$\mu$m down to 90\,nm.
\begin{figure*}[ht]
\centering
\includegraphics[width=\textwidth, angle=0]{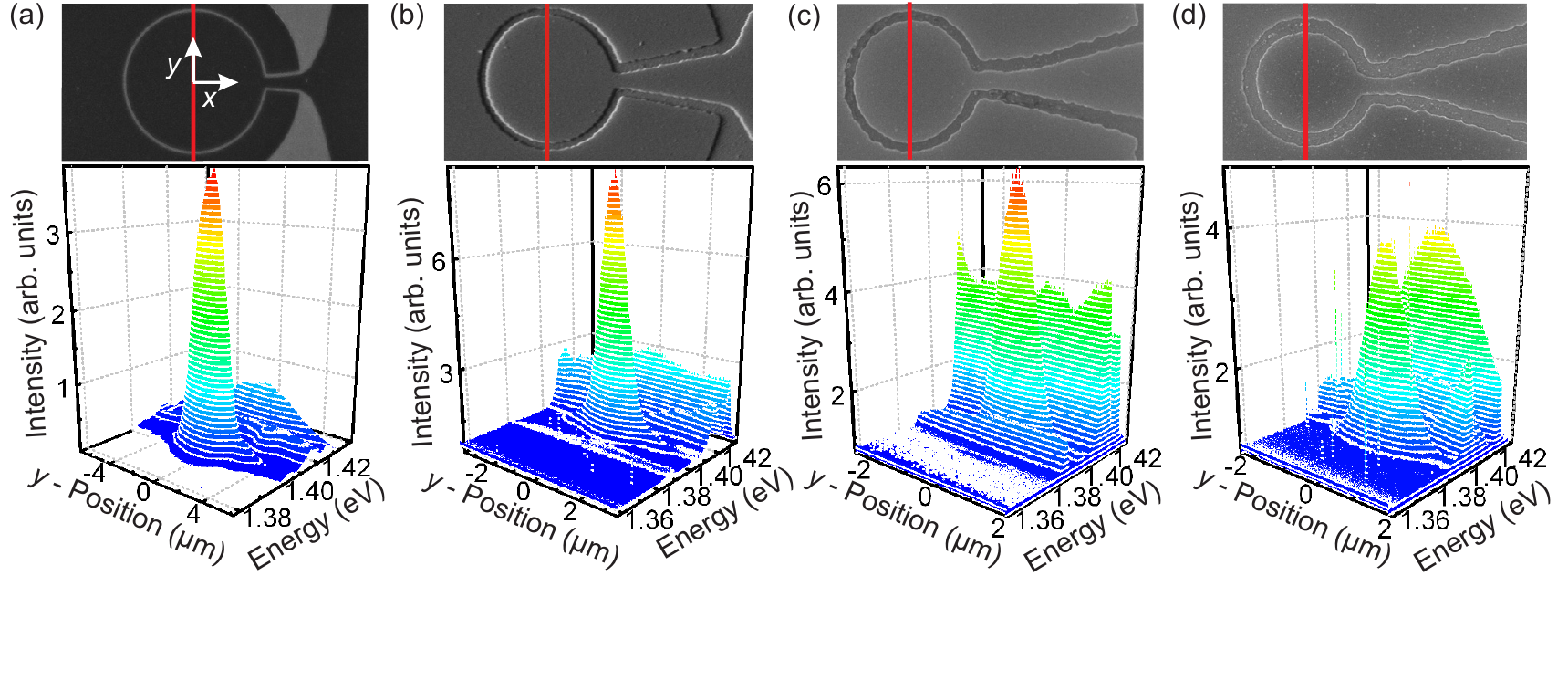}
\caption{\label{Circular traps} \textbf{Circular traps with different diameters.} The figure shows a series of scanning electron microscope (SEM) images of circular traps consisting of a central trap gate, with different diameters $d_{\mathrm{trap}}$, which are surrounded by a guard gate. In the lower row of pseudo-3D pictures, the photoluminescence intensity is shown energetically and spatially resolved. The cuts are taken along the $y$-direction through the middle of the trap as indicated by the red line in the SEM images. The exciton generation by a focused pump laser occurs a distance $x_{\mathrm{Laser}}$ left of the trap center below the guard gate. (a) [$d_{\mathrm{trap}}$ = 4000\,nm, $x_{\mathrm{Laser}}$ = -5\,$\mu$m, $V_{\mathrm{G}}$ = 0.60\,V, $V_{\mathrm{T}}$ = 0.35\,V, $T_{\mathrm{Lattice}}$ = 4\,K, $E_{\mathrm{Laser}}$ = 1.959\,eV, $P_{\mathrm{Laser}}$ = 10\,$\mu$W]; (b) [$d_{\mathrm{trap}}$ = 1800\,nm, $x_{\mathrm{Laser}}$ = -5\,$\mu$m, $V_{\mathrm{G}}$ = 0.55\,V, $V_{\mathrm{T}}$ = 0.20\,V, $T_{\mathrm{Lattice}}$ = 8\,K, $E_{\mathrm{Laser}}$ = 1.959\,eV, $P_{\mathrm{Laser}}$ = 2\,$\mu$W];  (c) [$d_{\mathrm{trap}}$ = 1000\,nm, $x_{\mathrm{Laser}}$ = -4\,$\mu$m, $V_{\mathrm{G}}$ = 0.55\,V, $V_{\mathrm{T}}$ = 0.20\,V, $T_{\mathrm{Lattice}}$ = 245\,mK, $E_{\mathrm{Laser}}$ = 1.959\,eV, $P_{\mathrm{Laser}}$ = 2\,$\mu$W]; (d) [$d_{\mathrm{trap}}$ = 800\,nm, $x_{\mathrm{Laser}}$ = -1.8\,$\mu$m, $V_{\mathrm{G}}$ = 0.55\,V, $V_{\mathrm{T}}$ = 0.20\,V, $T_{\mathrm{Lattice}}$ = 247\,mK, $E_{\mathrm{Laser}}$ = 1.494\,eV, $P_{\mathrm{Laser}}$ = 2\,$\mu$W].}
\end{figure*}
The gate layout of the circular traps is always comparable and is patterned by e-beam lithography. The central circular trap gate is surrounded by a guard gate which is separated from a trap gate by a 100\,nm narrow (slit) ungated region for insulation. Suitably biased by voltage $V_{\mathrm{G}}$ below the flat band voltage of $V_{\mathrm{FB}}$ it defines the outside surface potential to enable the formation of IXs and creates a depletion zone \cite{1998wiemann, 2011Hermelin} around the trap gate perimeter, thus providing the confinement for IX. Fig.\ \ref{Circular traps} shows a series of circular traps with diameters of 4000\,nm, 1800\,nm, 1000\,nm and 800\,nm. The semitransparent gates are 6\,nm thick evaporated Ti films. The traps are investigated in a fiber based low temperature microscope, composed of two diffraction limited confocal objectives, embedded in a $^{\mathrm{3}}$He refrigerator at variable temperatures down to 240\,mK \cite{2011Schinner-correlation}. The exciton generation by a focused pump laser occurs left of the trap on the guard gate. After a short relaxation and cool-down time of less than 1\,ns a large fraction of the equilibrated indirect exciton gas enters the energetically favorable trap area for $V_\mathrm{T} < V_\mathrm{G}$ \cite{2009Hammack}. The trapping potential is attractive for indirect excitons but repels unbound electrons \cite{2011Schinner}. The pseudo-3D pictures in Fig.\ \ref{Circular traps} display the measured PL intensity as a function of the PL energy and $y$-direction in a cut through the middle of the trap as indicated by the red line in the corresponding scanning electron microscope picture. The PL light is collected in transmission with a second confocal objective positioned below the sample. As reflected by the strong indirect exciton PL signal from the trapping area the trap provides a perfect confinement for indirect excitons.

\section{\label{ConfinementQD}Confinement potential of circular traps}

\begin{figure}[b]
\centering
\includegraphics[width=8.0cm, angle=0]{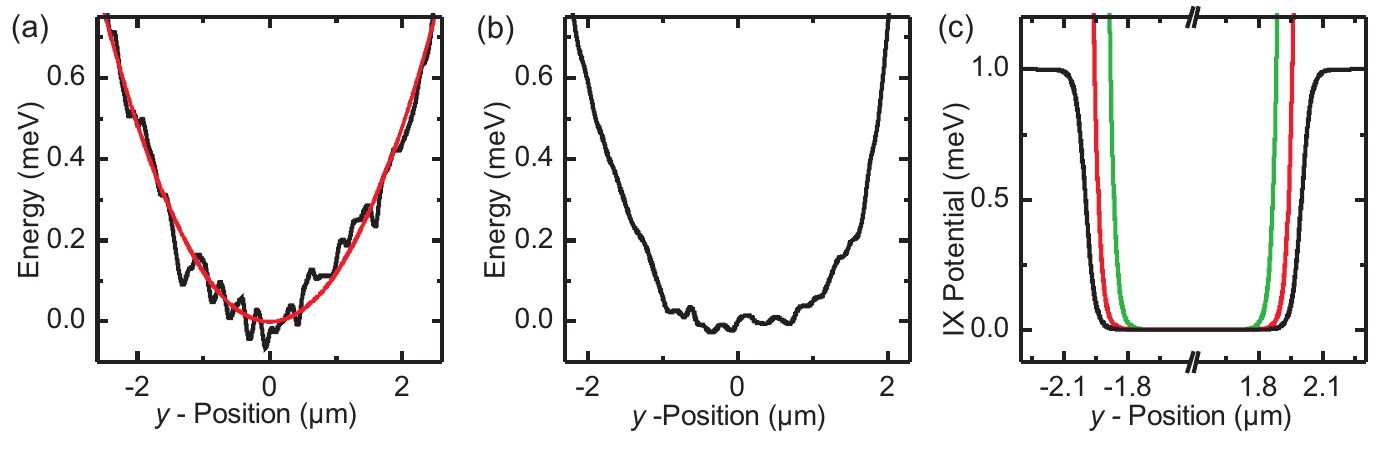}
\caption{\label{Confinement} \textbf{Confinement potential of a circular trap.} In (a) the measured confinement potential of a 25\,meV deep trap and a corresponding parabolic fit of a circular trap with a diameter of 4000\,nm (see Fig.\ \ref{Circular traps}(a)) is plotted [$d_{\mathrm{trap}}$ = 4000\,nm, $x_{\mathrm{Laser}}$ = -5\,$\mu$m, $V_{\mathrm{G}}$ = 0.60\,V, $V_{\mathrm{T}}$ = 0.20\,V, $T_{\mathrm{Lattice}}$ = 4\,K, $E_{\mathrm{Laser}}$ = 1.959\,eV, $P_{\mathrm{Laser}}$ = 9\,$\mu$W]. (b) shows the measured confinement potential of the same trap with 6\,meV energy difference between guard and trap exciton potential [$V_{\mathrm{G}}$ = 0.60\,V, $V_{\mathrm{T}}$ = 0.50\,V and all other parameters like (a)]. (c) shows the calculated confinement potential of a comparable trap with a deepness of 1\,meV (black), 8\,meV (red) and 40\,meV (green).}
\end{figure}

To characterize the confinement potential for IXs in a typical trap, we first discuss spatially resolved images of the PL energy of the trap with $d_{\mathrm{trap}}$ = 4000\,nm (Fig.\ \ref{Confinement}(a)), a diameter well above the resolution limit of our confocal optics which is comparable to the PL wavelength. In the case of a deep trap with a confinement of 25\,meV caused by a low trap gate voltage a parabolic confinement potential is observed (Fig.\ \ref{Confinement}(a)). For a shallow trap (6\,meV), with a relatively high trap gate voltage and a small voltage difference between the guard and trap gate, a box-like trapping potential is measured (Fig.\ \ref{Confinement}(b)), as predicted by simple electrostatic calculations (Fig.\ \ref{Confinement}(c)). For circular traps with a diameter exceeding 4000\,nm a box-like indirect exciton potential with flat bottom is observed as expected \cite{2009HighDiamond, 2011Schinner}. The potential landscape resulting from a given gate layout is strongly influenced by the thickness and the position of the coupled double quantum well (DQW) plane inside the field effect device \cite{2005Rapaport}. We positioned the DQW in the middle of the field-effect device. A position close to the heterostructure surface causes strong lateral electric fields in the quantum well plane resulting in exciton ionization on the trap perimeter. With increasing distance from surface gates, defining the trap geometry, the in-plane potential is blurred because of the exponential decay of the Fourier components \cite{1988Laux}. Solving the Laplace equation for a 4000\,nm diameter circular trap depicts a box-like excitonic potential (Fig.\ \ref{Confinement}(c)). In deeper traps an increasing depletion around the trap perimeter is calculated (Fig.\ \ref{Confinement}(c)). The boundary condition for these calculations is a homogenous dielectric medium between a metallic infinite back-contact and the top gate structure.
The real excitonic potential plotted in Fig.\ \ref{Confinement}(a) shows a stronger depletion, as theoretically expected, resulting in a parabolic confinement potential in the case of a deep trap. The reason for the differences to the calculated trapping potential is that the assumptions for the numeric solution disregard that we have a semiconductor layer system with an unknown background doping and a doped semiconductor back-contact containing a finite charge carrier density. In the case of the back-contact, it is expected that the applied electric field causes a depletion.
We point out that the depletion around the perimeter of the trap gate is the reason that no leakage of indirect excitons at the trap gate lead is observed. In two samples with a circular trap having a diameter of 600\,nm we measured a comparable quantum dot confinement. We further find experimentally that circular traps with diameters equal and smaller than 400\,nm are not able to trap excitons within the studied gate-voltage regime.

\section{\label{populatedQD}The quantum dot trap populated with many excitons}

\begin{figure}[b]
\centering
\includegraphics[width=8.6cm, angle=0]{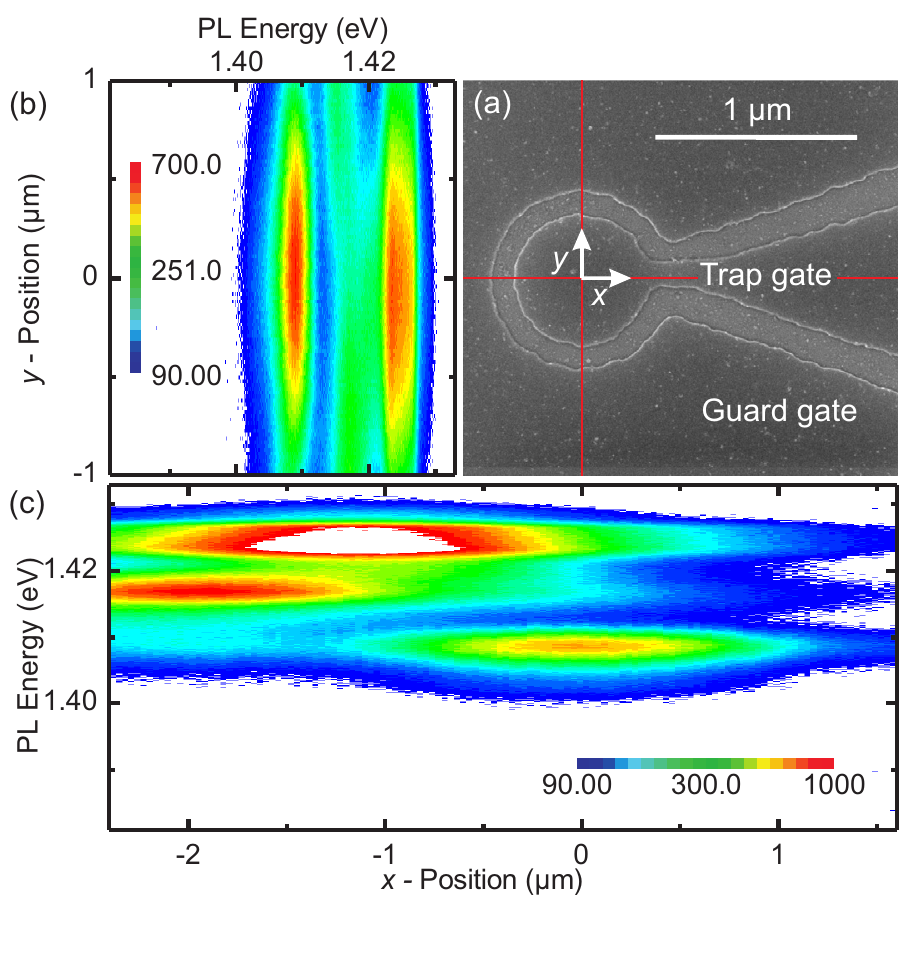}
\caption{\label{x-y-cuts}\textbf{The figure shows spatially and energy resolved measurements of the exciton distribution in the quantum dot trap populated with many excitons.} In (b) and (c) cuts are shown of the PL intensity in a logarithmic color scale, given by the scale bar, as a function of the position and PL energy. The data are measured along cuts indicated by the red lines in the scanning electron micrograph shown in (a) of the QD trap with a trap gate diameter of 600\,nm [$d_{\mathrm{trap}}$ = 600\,nm, $x_{\mathrm{Laser}}$ = -1.2\,$\mu$m, $V_{\mathrm{G}}$ = 0.65\,V, $V_{\mathrm{T}}$ = 0.40\,V, $T_{\mathrm{Lattice}}$ = 245\,mK, $E_{\mathrm{Laser}}$ = 1.494\,eV, $P_{\mathrm{Laser}}$ = 200\,nW].}
\end{figure}

An energetically deep circular trap with a trap gate diameter of 600\,nm (Fig.\ \ref{x-y-cuts}(a)) results in a quantum dot, for low exciton population. In Fig.\ \ref{x-y-cuts}(b), (c) the energetically and spatially resolved PL intensity is logarithmically color coded. The cuts are taken in $x$- and $y$-direction through the middle of the trap center. The exciton generation occurs by a strongly focused pump laser about 1.2\,$\mu$m away from the trap center below the guard gate. The energetically highest PL signal in Fig.\ \ref{x-y-cuts} is PL light emitted from direct quantum well excitons at $\sim$ 1.425\,eV. They have a short lifetime, recombining on a nanoseconds time scale \cite{1987Feldmann}. Hence, they exist only in the vicinity of the generation focus. The PL energetically below the direct exciton PL at $\sim$1.417\,eV results from indirect excitons under the guard gate. The energetically lowest PL light is emitted from long living IXs trapped in the high populated quantum dot at $\sim$1.409\,eV. The areal distribution of the trapped IX PL is spatially diffraction broadened to 1200\,nm full width at half maximum. This value is about 300\,nm broader than the spatial width observed for occupying the QD with a single IX (see Fig.\ \ref{QDfig1}(d)). We expect that the electrostatic trapping potential is modified by screening effects via the aligned dipole moments of the trapped indirect exciton ensemble. The consequence is a broadening of the trapping potential, and the population of the QD with many excitons results in a strong blue shift leading to a more shallow trapping potential.

\begin{figure}[b]
\centering
\includegraphics[width=6cm, angle=0]{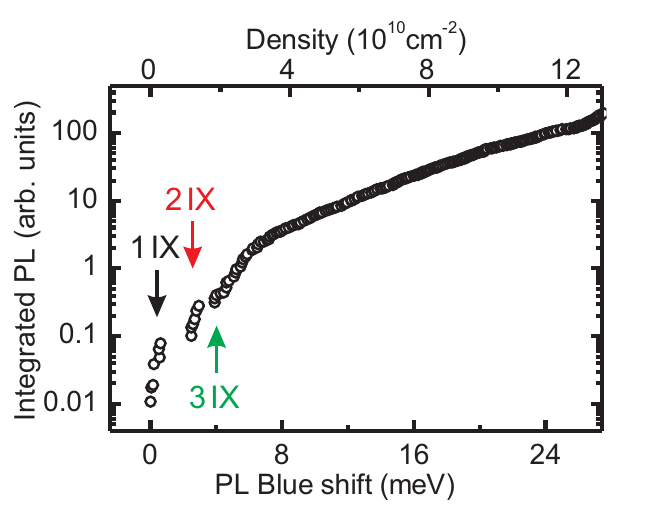}
\caption{\label{QDdensity} \textbf{Emitted PL intensity versus PL blue shift.} The integrated PL intensity is logarithmically plotted as a function of the PL blue shift and the corresponding calculated trapped indirect exciton density [$d_{\mathrm{trap}}$ = 600\,nm, $x_{\mathrm{Laser}}$ = -1.2\,$\mu$m, $V_{\mathrm{G}}$ = 0.65\,V, $V_{\mathrm{T}}$ = 0.30\,V, $T_{\mathrm{Lattice}}$ = 247\,mK, $P_{\mathrm{Laser}}$ = 150\,pW...850\,nW, $E_{\mathrm{Laser}}$ = 1.494\,eV].}
\end{figure}

A change of the QD population can easily be studied by changing the excitation laser power. In Fig.\ \ref{QDdensity} the overall integrated trap PL intensity is displayed logarithmically as a function of the PL blue shift $\Delta E$ as well as a rough estimate of the corresponding exciton density as detailed below. The blue shift $\Delta E$ is measured relative to the PL energy of a single indirect exciton. For higher populations of the QD, the maximum PL energy of the strongest emitting PL line is used to calculate the blue shift belonging to the total intensity detected from the QD. The lowest PL intensity is emitted from a single 1\,IX which experiences only the electric field created by the gates and causing the confinement potential. A biexciton 2\,IX in the QD produces a step like increase in the PL maximum energy by about 2\,meV via dipolar repulsion (Fig.\ \ref{QDdensity} or Fig.\ \ref{QDfig12}(c)). The dipole moments of all trapped individual IXs are aligned and perpendicular to the QW-plane. From the dipole interaction energy of 2\,meV, a lateral interexcitonic distance $r_{\mathrm{IX}}$ = 26\,nm is calculated by using $W=\frac{p^2}{\epsilon r_{\mathrm{IX}}^{3}}$ where $\mathbf{p}=e\mathbf{d}$ is the IX dipole moment resulting from the electron-hole separation by a distance $\mathbf{d}$ in the two adjacent quantum wells and $\epsilon$ is the dielectric constant. $r_{\mathrm{IX}}$ is slightly larger than the effective Bohr radius $r_{\mathrm{e}}=21.7$\,nm which is extracted from the diamagnetic energy shift by a quadratic fit (see Fig.\ \ref{QDfig2}(f)).
For a QD populated with a triexciton 3\,IX a blue shift of 3\,meV is measured (Fig.\ \ref{QDdensity} or Fig.\ \ref{QDfig12}(c)). Taking the assumption that all three excitons are equidistantly spaced an interexcitonic distance of $r_{\mathrm{IX}}$ = 28\,nm is calculated. And for 3\,IX an effective Bohr radius of $r_{\mathrm{e}}=21.7$\,nm is found (Fig.\ \ref{QDfig2}(f)).

The strong dipolar interactions between the trapped IXs in the quantum dot can cause spatial order of the excitons, in analogy to a Wigner-crystal.
This spatial order is likely to be assisted by the fact that the external excitonic potential has a significantly larger spatial extent than the excitons confined to the Bohr radii. In the case that the electron wavefunctions show a strong overlap for $r_{\mathrm{e}}> r_{\mathrm{IX}}/2 $, one can anticipate to observe correlations because the expected excitonic coherence length \cite{2006Yang} exceeds the diameter of the trap. Furthermore, the thermal de Broglie wavelength $\lambda_{\mathrm{dB}}$ of the bosonic particles is larger than $r_{\mathrm{IX}}$ which is similar to the effective Bohr radius $r_{\mathrm{e}}$.

For higher exciton populations of the QD, the effective electric field, seen by the individual dipolar indirect exciton is the sum of the externally applied electric field and the depolarizing dipole field of all other IXs in the vicinity. With increasing exciton density, the effective electric field decreases due to screening and causes a blue shift $\Delta E$. From $\Delta E$ it is possible to calculate the corresponding exciton density. The simplest approach is that the indirect exciton density is given by $n = \frac{\epsilon }{4\pi e^{2}d }\Delta E$ \cite{2011Schinner}. This equation likely underestimates the density, especially in the low density limit as theoretically calculated by \cite{2007Zimmermann, 2008Schindler, 2010Ivanov}. In the special case of the QD, however, we used this formula to convert the blue shift in the density given on top of Fig.\ \ref{QDdensity}.

We expect that the electrostatic trapping potential is modified by the screening effects caused by the aligned dipole moments of the trapped indirect exciton ensemble. The consequence is a broadening of the trapping potential and a more shallow confinement potential. In the case of a high exciton density in the quantum dot (Fig.\ \ref{x-y-cuts}), a populated trap area with a diameter of about 300\,nm is estimated. Electrostatic calculations show a trap diameter of 350\,nm to 400\,nm. From these values we calculate a maximum QD population of approximately 90 up to 150 trapped indirect excitons at high laser powers.

In Fig.\ \ref{QDdensity} the trapped exciton number increases by a factor of $10^2$. In contrast, the emitted PL intensity changes by more than a factor $10^4$ and it shows two different slopes. A part of this exponential increasing intensity as a function of the raising trapped exciton number can be explained with a decreasing lifetime \cite{2011Schinner}. The rest is associated with a collective behavior of the exciton ensemble and we believe that correlations in the bosonic exciton gas start to be important.

\section{Asymmetry of the lineshape at large exciton occupation}

At the low temperatures of our experiment the kinetic energy of the excitons is negligible in comparison to the typical energies of interexcitonic dipolar repulsion \cite{1990Yoshioka, 2009Laikhtman}. At $T=$0\,K this repulsion causes the exciton state with the highest occupation $N$ to have the highest transition energy and strongest oscillator strength and defines the blue edge of the total IX emission. States of $N-M$ excitons with $M=1,2, \ldots N$ excitons occur with increasing $M$ at successively lower energies but still contribute to the emission spectrum via the recombination cascade.  Assuming constant dipolar repulsion this results in a two-dimensional exciton systems with a joint optical density of states proportional to energy  $E_{N}-E_{0}$. Furthermore, one can expect the radiative lifetime of the $N-M$ state to rise with increasing $M$, thus causing a strongly asymmetric $T=0$ lineshape. Additional inhomogeneous broadening caused by temporal fluctuation of the electrostatic background will smear the blue edge and merge the energetically lower lying transitions in the recombination cascade without much affecting the red tail of the PL lineshape.

\section{\label{ModelQD}Modeling the energy spectra in a few-exciton trap}

\begin{figure}[b]
\centering
\includegraphics[width=8.6cm, angle=0]{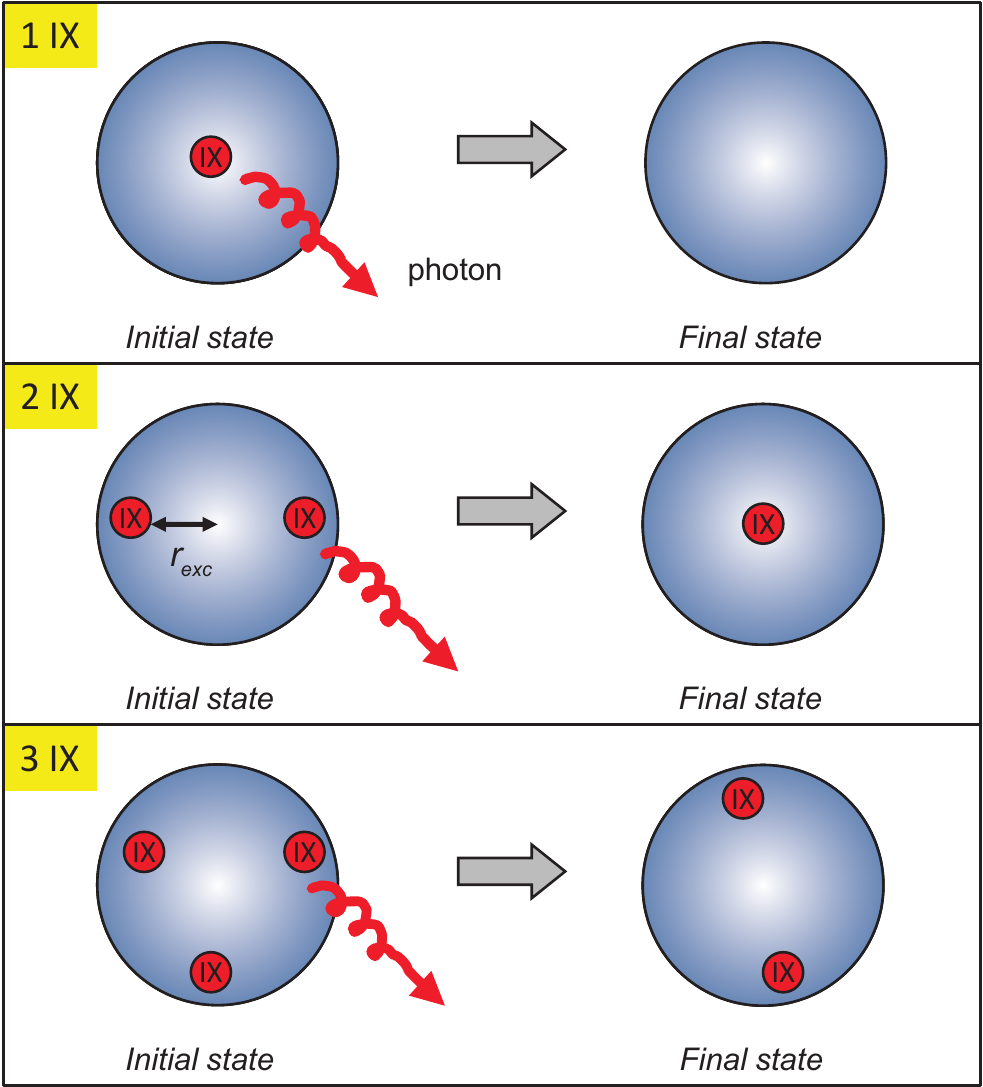}
\caption{\label{Model1} \textbf{Wigner-molecule representation of exciton complexes.} An emission process brings a multiple exciton (2\,IX, 3\,IX, $\dots$) into a long-lived, fully-relaxed ground state.}
\end{figure}

We now estimate optical energies of few-exciton states in a trap by using the Wigner-crystal-like model (Fig.\ \ref{Model1}). Such a model has been applied to few-electron states in electron quantum dots \cite{1994Peeters, 1996Govorov} and is extended here to parallel aligned and spatially ordered dipolar excitons \cite{1990Yoshioka, 2009Laikhtman}. In this quasi-classical approach, in which overlap of the excitonic wavefunctions is neglected, the total exciton energy in a trap is written in the following way:
\begin{equation}
\label{MEq1}
E_{\mathrm{exc},\,N_{\mathrm{exc}}} =N_{\mathrm{exc}}E^{0}_{\mathrm{IX}}+\sum_{i}\frac{M\Omega^{2}_{\mathrm{exc}}}{2}r_{i}^{2}+\sum_{i<j\atop i,j=1\dots N_{\mathrm{exc}}}\frac{p^{2}}{\epsilon r_{ij}^{3}},
\end{equation}
where $N_{\mathrm{exc}}$ is the number of excitons in a trap, $E^{0}_{\mathrm{IX}}$ is the optical energy of the exciton in an uniform 2D well, $\Omega_{\mathrm{exc}}$ is the characteristic confinement frequency of an exciton in a parabolic trap, $M=m^{*}_{\mathrm{e}}+m^{*}_{\mathrm{hh}}$ is the exciton mass, $p$ is the exciton dipole, and $r_{i}$ is a coordinate of $i$-exciton in a trap. Then, the ground state should be found as a minimum of the total energy $E_{\mathrm{exc}, N_{\mathrm{exc}}}$. Since excitons have enough time to relax, this ground state should be considered and it is the initial state for the photon emission process (Fig.\ \ref{Model1}). For the first three excitonic complexes ($N_{\mathrm{exc}}=1-3$), the ground states are shown in Fig.\ \ref{Model1} and they can easily be calculated from Eq.\ \ref{MEq1}. The equilibrium positions of correlated excitons are determined by the interplay of the parabolic confining potential and the dipole repulsion. The emission process can be considered as a fast removal of one of the correlated excitons and ends in a final state of $(N_{\mathrm{exc}}-1)$\,IX. Now we calculate the emission energy of $N_{\mathrm{exc}}$\,IX which is an energy difference
\begin{equation}
\label{MEq2}
\hbar \omega_{N_{\mathrm{exc}}}=E_{\mathrm{initial},\,N_{\mathrm{exc}}}(r_{i,\,N_{\mathrm{exc}}})-E_{\mathrm{final},\,N_{\mathrm{exc}}-1}(r_{i,\,N_{\mathrm{exc}}-1}),
\end{equation}
where $E_{\mathrm{initial},\,N_{\mathrm{exc}}}$ is the energy found as the minimum of Eq.\ \ref{MEq1} and $r_{i,\,N_{\mathrm{exc}}}$ are the equilibrium positions of the excitons in the initial state with a minimized energy. $E_{\mathrm{final},\,N_{\mathrm{exc}}-1}$ is the total energy of the final state calculated as a minimum of Eq.\ \ref{MEq1} for the system of $(N_{\mathrm{exc}}-1)$ excitons with positions $r_{i,\,N_{\mathrm{exc}}-1}$. Figure \ref{Model1} illustrates the physical situation.

The initial state of 1\,IX is a single exciton localized in the center of trap, whereas the final state of 1\,IX after the emission process is, of course, a band structure ground state. This case is really simple and its emission energy is: $\hbar \omega_{N_{\mathrm{exc}}=1}=E^{0}_{\mathrm{IX}}$. For 2\,IX, we should first minimize the total energy and find the equilibrium positions of the excitons. Then, we can find the emission energy. The calculations along these lines yield:
\begin{equation}
r_{\mathrm{exc},\,N_{\mathrm{exc}}=2}=\sqrt[5]{\frac{3}{16\epsilon}\frac{p^{2}}{M\om}},
\end{equation}
$E_{\mathrm{initial},\,N_{\mathrm{exc}}=2}(r_{i})=E_{\mathrm{exc},\,N_{\mathrm{exc}}=2}=2E^{0}_{\mathrm{IX}}$
\begin{equation}
+2\frac{M\om}{2}r^{2}_{\mathrm{exc},\,N_{\mathrm{exc}}=2}+\frac{p^{2}}{8\epsilon r^{3}_{\mathrm{exc},\,N_{\mathrm{exc}}=2}}.
\end{equation}
The final state energy is
\begin{equation}
E_{\mathrm{final},\,N_{\mathrm{exc}}=1}(r_{i})=E^{0}_{\mathrm{IX}}.
\end{equation}
Then, the optical emission energy becomes:
\begin{equation}
\hbar \omega_{N_{\mathrm{exc}}=2}=E^{0}_{\mathrm{IX}}+2\frac{M\om}{2}r^{2}_{\mathrm{exc},\,N_{\mathrm{exc}}=2}+\frac{p^{2}}{8\epsilon r^{3}_{\mathrm{exc},\,N_{\mathrm{exc}}=2}}.
\end{equation}
For 3\,IX, the ground-state configuration is of triangular geometry. The corresponding parameters of the exciton complex are
\begin{equation}
E_{\mathrm{exc},\,N_{\mathrm{exc}}=3}=3E^{0}_{\mathrm{IX}}+3\frac{M\om}{2}r^{2}_{\mathrm{exc},\,N_{\mathrm{exc}}=3}+3\frac{p^{2}}{\alpha^{3}\epsilon r^{3}_{\mathrm{exc},\,N_{\mathrm{exc}}=3}},
\end{equation}
\begin{equation}
r_{\mathrm{exc},\,N_{\mathrm{exc}}=3}=\sqrt[5]{\frac{3}{\alpha^{3}\epsilon}\frac{p^{2}}{M\om}},
\end{equation}
where $\alpha=\sqrt{3}$. The optical energy becomes: \\
$\hbar \omega_{N_{\mathrm{exc}}=3} =E_{\mathrm{exc},\,N_{\mathrm{exc}}=3}-E_{\mathrm{exc},\,N_{\mathrm{exc}}=2} =\\
=E^{0}_{\mathrm{IX}}+\frac{M\om}{2}(3r^{2}_{\mathrm{exc},\,N_{\mathrm{exc}}=3}-2r^{2}_{\mathrm{exc},\,N_{\mathrm{exc}}=2})+$
\begin{equation}
+3\frac{p^{2}}{\alpha^{3}\epsilon r^{3}_{\mathrm{exc},\,N_{\mathrm{exc}}=3}}-\frac{p^{2}}{8\epsilon r^{3}_{\mathrm{exc},\,N_{\mathrm{exc}}=2}}.
\end{equation}
We now can calculate the optical energies for the three complexes. Using the overall realistic parameters, $\Omega_{\mathrm{exc}}$ = 0.8\,meV and $M=0.5\,m_{0}$, $\frac{p}{|e|}$ = 17\,nm, and $\epsilon$ = 12, we can obtain a reasonable agreement with the experimental data for the emission spectrum:\\
$\hbar \omega_{N_{\mathrm{exc}}=1}=E^{0}_{\mathrm{IX}} \\
\hbar \omega_{N_{\mathrm{exc}}=2}=E^{0}_{\mathrm{IX}} + 2.1\,$meV \\
$\hbar \omega_{N_{\mathrm{exc}}=3}=E^{0}_{\mathrm{IX}} + 2.8\,$meV.\\
We note that the only fitting parameter here was the confining potential, $\Omega_{\mathrm{exc}}$ = 0.8\,meV. The corresponding radii of exciton complexes are:\\
$r_{\mathrm{exc},\,N_{\mathrm{exc}}=1}=0\\
r_{\mathrm{exc},\,N_{\mathrm{exc}}=2}=34.5$\,nm\\
$r_{\mathrm{exc},\,N_{\mathrm{exc}}=3}=37.5$\,nm.\\
\begin{figure}[t]
\centering
\includegraphics[width=8.6cm, angle=0]{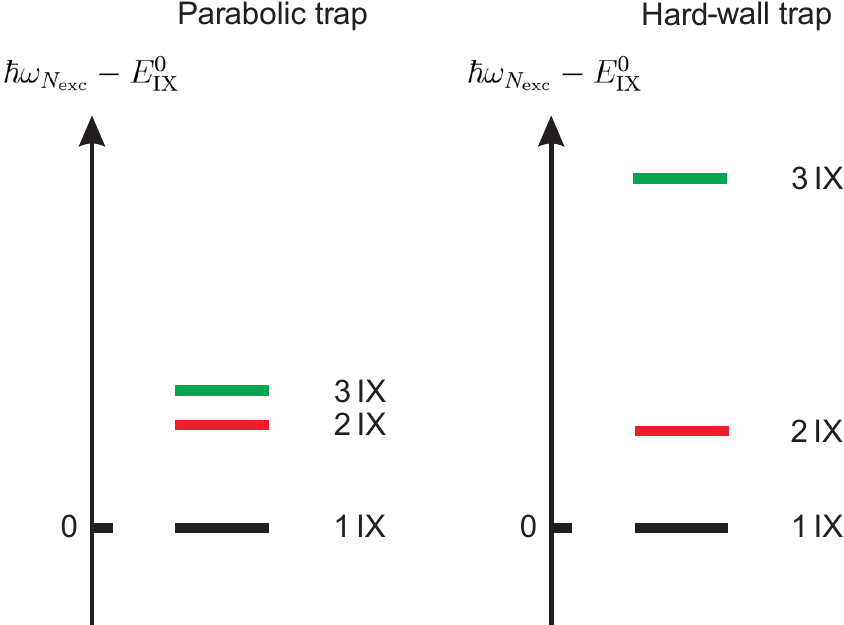}
\caption{\label{Model2} \textbf{Quantum dot level structure.} Qualitative diagrams for the spectra of excitons, $N_{\mathrm{exc}}$\,IX, confined in the parabolic and hard-wall traps.}
\end{figure}%
The above numbers exceed an estimated Bohr radius of an exciton thus validating our model. We see that the energy spacing between the consequent excitonic states decreases (Fig.\ \ref{Model2}). This is a property of a parabolic trap. For a hard-wall trap, we obtain a qualitatively different spectrum. In a hard-wall trap due to the Coulomb repulsion, the radial coordinates of excitons for 2\,IX and 3\,IX are fixed by the hard walls, $r_{i}=R_{\mathrm{trap}}$. The excitons tend to be as far as possible from each other and, therefore, they are located at the walls. The ground state configurations are a diatomic molecule and a triangle (like in Fig.\ \ref{Model1}). Then, the optical energies are calculated as:\\
$\hbar \omega_{N_{\mathrm{exc}}=1}-E^{0}_{\mathrm{IX}}=0 \\
\hbar \omega_{N_{\mathrm{exc}}=2}-E^{0}_{\mathrm{IX}} =\frac{p^{2}}{8\epsilon R_{\mathrm{trap}}^{3}}\approx1.97\,$meV \\
$\hbar \omega_{N_{\mathrm{exc}}=3}-E^{0}_{\mathrm{IX}} =2\frac{p^{2}}{\alpha^{3}\epsilon R_{\mathrm{trap}}^{3}}\approx7.1\,$meV\\
assuming the trap diameter $2R_{\mathrm{trap}}$ = 26\,nm. We see here a striking difference to the case of parabolic trap. The energy intervals for the consequent exciton lines increase with the number of excitons (Fig.\ \ref{Model2}). In a hard-wall trap, the Coulomb repulsion energy builds up with increase of number of excitons since the exciton system should occupy a fixed area. Therefore, the inter-particle distance decreases and Coulomb correlations grow. As a result, the energy interval between the excitonic peaks in a spectrum increases with $N_{\mathrm{exc}}$. In contrast, a parabolic trap has "soft" walls and the size of an excitonic complex increases with the number of excitons in a trap. Therefore, the increase in the correlation energy is somewhat smaller after addition of an exciton. This results in a different multi-exciton spectrum in which an energy interval decreases with increasing the exciton number (Fig.\ \ref{Model2}).  A qualitative agreement between the theory and the experiment gives us another argument that the traps in our experimental samples are parabolic-like.

\section{\label{LifetimeQD} Lifetime of trapped indirect excitons}

\begin{figure}[b]
\centering
\includegraphics[width=8.6cm, angle=0]{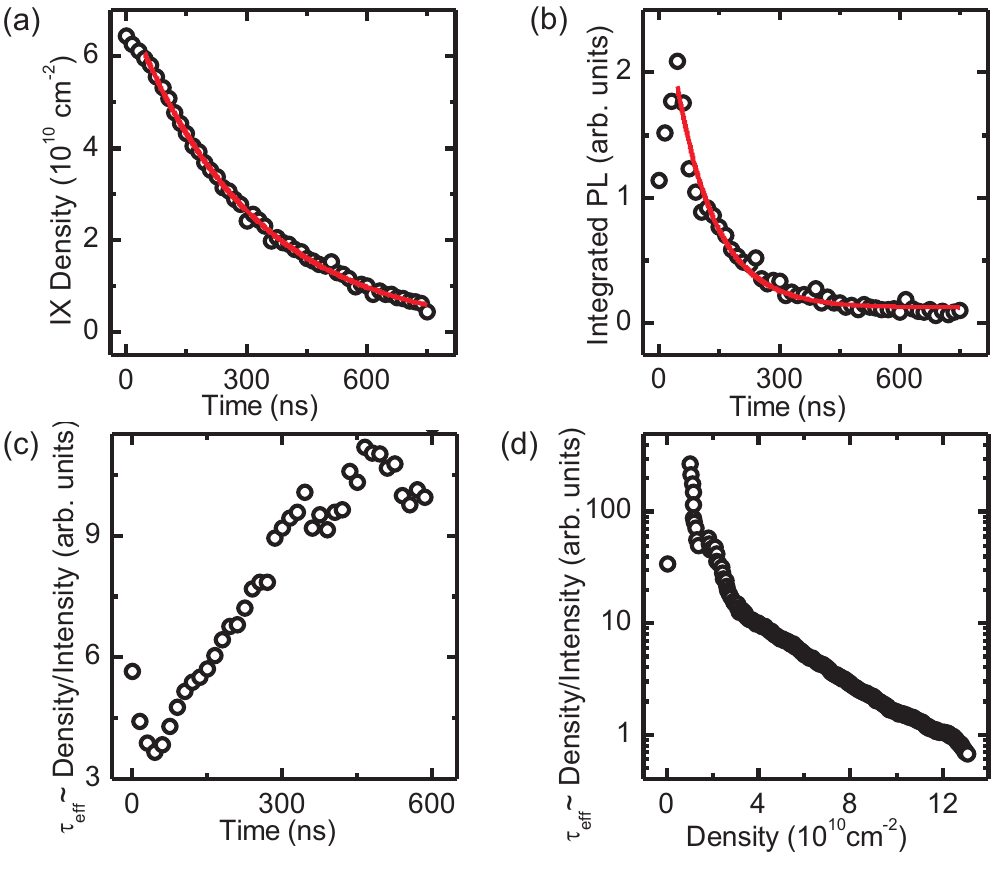}
\caption{\label{time} \textbf{Time resolved measurement of the exciton population.} It is plotted the decay of the exciton density in (a), and in (b) the decay of the integrated PL intensity, as a function of time after a 40\,ns excitation laser pulse. Red lines are mono exponential decay fits. In (c) the effective exciton lifetime extracted from the data shown in (a) and (b) as a function of time is plotted. The effective exciton lifetime, extracted from the data shown in Fig.\ \ref{QDdensity}, is plotted as a function of the trapped exciton density in (d) [$d_{\mathrm{trap}}$ = 600\,nm, $x_{\mathrm{Laser}}$ = -1.2\,$\mu$m, repetition rate 500\,kHz, $V_{\mathrm{G}}$ = 0.65\,V, $V_{\mathrm{T}}$ = 0.25\,V, $T_{\mathrm{Lattice}}$ = 250\,mK, $P_{\mathrm{Laser}}^{\mathrm{cw}}$ = 1.8\,$\mu$W, $E_{\mathrm{Laser}}$ = 1.494\,eV].}
\end{figure}

The lifetime of an indirect exciton depends strongly on the overlap of the tails of the electron and hole wavefunction forming the IX. We explore the temporal dynamics of the trapped indirect exciton ensemble in time resolved PL spectra. Replacing the continuous-wave laser by a pulsed laser with the same emission wavelength, we excite excitons in the vicinity of the trap under the guard gate as in Fig.\ \ref{x-y-cuts}. From the time-resolved PL spectra the PL maximum energy is extracted in order to calculate the trapped exciton density, as a function of delay time, by using the simple equation mentioned above with $n(t) \sim \Delta E(t)$ (Fig.\ \ref{time}(a)).
The density shows an exponential decay with a time constant $\tau_\mathrm{n}=(302\pm7)$\,ns. The corresponding PL intensity in Fig.\ \ref{time}(b) shows a decay constant of  $\tau_\mathrm{I}=(99\pm5)$\,ns, a factor 3 different from $\tau_\mathrm{n}$. In the time-resolved measurements, our sensitivity is too low to detect the few-exciton regime and therefore we expect that the real exciton density is 30\% higher.
The long IX lifetimes also demonstrate that even in the narrow confinement potential exciton ionization is rather unlikely.
One reason for the different time constants is that with decreasing density, corresponding to  $\Delta E(t)$, the indirect exciton lifetime increases. According to Ref.\ \cite{2011Schinner}, it is possible to define a time-dependent effective lifetime $\tau_\mathrm{eff}(t) \sim n(t)/I(t)$ as the ratio of density and emitted intensity. In Fig.\ \ref{time}(c) this effective lifetime increases with raising delay time and accordingly with decreasing trapped exciton density. The effective lifetime as a function of the exciton density is shown in Fig.\ \ref{time}(d). To extract $\tau_\mathrm{eff}(n)$ the continuous-wave excitation data illustrated in Fig.\ \ref{QDdensity} are used. With increasing exciton density the ratio of density and intensity and hence $\tau_\mathrm{eff}(n)$ decreases over more than two orders of magnitude.
From the result that $\tau_\mathrm{eff}$ increases with decreasing exciton densities (Fig.\ \ref{time}(c), (d)) we anticipate that for a few trapped IXs the lifetime is much longer than the observed one.
An increasing exciton density (Fig.\ \ref{time}(d)) is accompanied by a strong blue shift caused by the screening of the external applied electrical field. This reduction in the effective Stark field causes a strongly increasing overlap of the tails of the electron and hole wavefunctions forming the IX thus increasing the oscillator strength, and decreasing the lifetime $\tau_\mathrm{eff}(n)$. In addition, one can expect that at high exciton densities, collective properties and correlated behavior, induced by the low temperatures (240\,mK), modify the emission properties of the trapped excitonic ensemble similar to the nonlinear
gain observed in exciton-polariton condensates \cite{2007Balili}.

\section{Temperature dependence}

\begin{figure}[ht]
\centering
\includegraphics[width=4cm, angle=0]{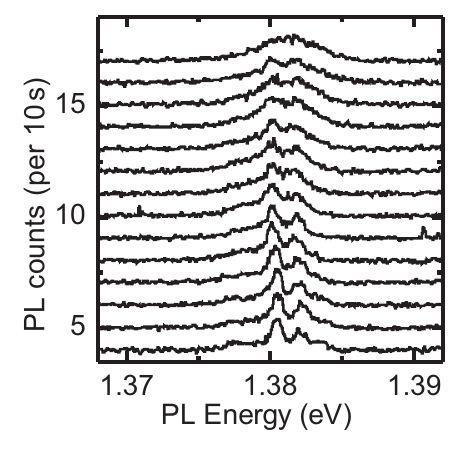}
\caption{\label{QDtemperature} \textbf{Temperature dependence of the quantum dot photoluminescence.} Energy resolved PL spectra for different temperatures in equidistant steps from 250\,mK (bottom) to 6.75\,K (top) are shown [$V_{\mathrm{G}}$ = 0.55\,V, $V_{\mathrm{T}}$ = 0.25\,V, $P_{\mathrm{Laser}}$ = 3\,nW, $E_{\mathrm{Laser}}$ = 1.494\,eV].}
\end{figure}

Comparable to gate defined charge quantum dots with a small charging energy, the quantum dot characteristics are only visible at low temperatures. In Fig.\ \ref{QDtemperature} the temperature dependence of the photoluminescence spectra emitted from the quantum dot is plotted. With increasing temperature, the discrete PL lines broaden and merge, resulting in an unstructured somewhat asymmetric PL lineshape. At elevated temperatures, thermally activated scattering processes are expected. Possible interactions are IX-phonon scattering, IX-IX scattering and shake up processes, causing the broadened spectra. In addition, as the thermal energy gets comparable to the characteristic spatial quantization energy of the IX, the IX become increasingly delocalized in the trap. The Wigner-molecule unbinds, causing a broadening of the PL lineshape, particularly on the high energy side.

\section{\label{Leakage currents}Leakage currents}

\begin{figure}[h]
\centering
\includegraphics[width=4cm, angle=0]{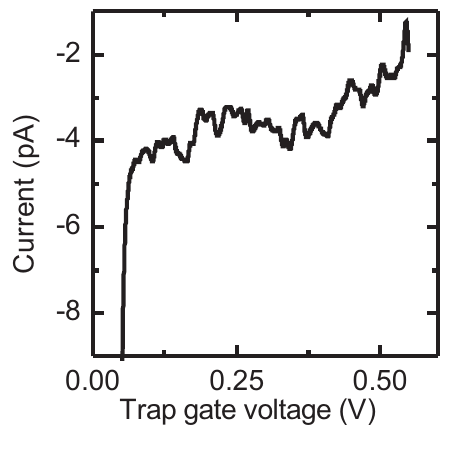}
\caption{\label{current} \textbf{The current flow on the trap gate.}  The current flow from the trap gate to the back gate is plotted versus trap gate voltage [$V_{\mathrm{G}}$ = 0.65\,V, $T_{\mathrm{Lattice}}$ = 244\,mK, $P_{\mathrm{Laser}}$ = 2.8\,nW, $E_{\mathrm{Laser}}$ = 1.494\,eV]}
\end{figure}

With decreasing trap gate voltage we observe an unexpected kink in the slope of the quantum-confined Stark effect (see Fig.\ \ref{QDfig2}(c)). Therefore, the current flow from the trap gate to the back gate is measured to exclude leakage currents as the origin (Fig.\ \ref{current}). However, leakage currents entering the trap gate are negligibly small in the corresponding voltage regime where the PL is emitted. The absolute value of the current has an unknown constant offset of only a few pA. Thus, it is reasonable to assume that the externally applied voltages fully reflect the potentials applied at the gates in the relevant range of trap gate voltages. The kink in the slope of the quantum-confined Stark effect is expected to be a consequence of the narrowing of the quantum dot potential with increasing trap depth. In turn, the PL line of both a 1\,IX and a 2\,IX are detected. Only for a trap gate voltage smaller than 0.06\,V, a significant current starts to flow through the diode structure (Fig.\ \ref{current}).

\section{Conclusion}

In conclusion, we realized individual optically active, gate-defined and voltage-controlled quantum traps for spatially indirect excitons in which a tuneable potential causes excitonic confinement in the DQW-plane. The population of such a QD can be tuned from a single exciton to a multi exciton state of above 100\,IXs. With our device we introduce a gate-defined platform for creating and controlling optically active quantum dots with possible applications in fundamental quantum physics and quantum computing. This provides the possibility of lithographically defined coupled quantum dot arrays with voltage-controlled optical properties and tuneable in-plane coupling.

\textbf{Acknowledgements:} We thank M.\,P.\,Stallhofer, D.\,Taubert, S.\,Ludwig, K.\,Kowalik-Seidl and A.\,H\"ogele for stimulating discussions and helpful comments. Financial support by the Deutsche Forschungsgemeinschaft under Project No. Ko 416/17, the SPP1285 as well as the German Excellence Initiative via the Nanosystems Initiative Munich (NIM), LMUexcellent and BMBF QuaHL-Rep (01 BQ 1035) is gratefully acknowledged.

\end{document}